\documentclass[fleqn,10pt]{wlscirep}

\usepackage{color,amsmath,amssymb,graphicx,epstopdf,bm,soul}

\title{Multivalley engineering in semiconductor microcavities}

\author[1*]{M. Sun}
\author[1,2,3]{I. G. Savenko}
\author[4]{H. Flayac}
\author[5]{T. C. H. Liew}

\affil[1]{Center for Theoretical Physics of Complex Systems, Institute for Basic Science, Daejeon, Republic of Korea}
\affil[2]{Nonlinear Physics Centre, Research School of Physics and Engineering, The Australian National University, Canberra ACT 2601, Australia}
\affil[3]{ITMO University, St. Petersburg 197101, Russia}
\affil[4]{Institute of Theoretical Physics, Ecole Polytechnique F\'ed\'erale de Lausanne (EPFL), CH-1015 Lausanne, Switzerland}
\affil[5]{Division of Physics and Applied Physics, School of Physical and Mathematical Sciences, Nanyang Technological University, 21 Nanyang Link, Singapore 637371}

\affil[*]{sunmeg.89@gmail.com}

\begin{abstract}
We consider exciton-photon coupling in semiconductor microcavities in which separate periodic potentials have been embedded for excitons and photons. We show theoretically that this system supports degenerate ground-states appearing at non-zero in-plane momenta, corresponding to multiple valleys in reciprocal space, which are further separated in polarization corresponding to a polarization-valley coupling in the system. Aside forming a basis for valleytronics, the multivalley dispersion is predicted to allow for spontaneous momentum symmetry breaking and two-mode squeezing under non-resonant and resonant excitation, respectively.
\end{abstract}
\begin{document}

\flushbottom
\maketitle

\section*{Introduction}

Photonic and electronic systems support many common universal phenomena. To give a selection of examples, the field of topological photonics emerged recently from ideas in the study of topological insulators~\cite{Lu2014}, and the field of spintronics has been an inspiration for optical analogues in the optical spin Hall effect~\cite{Leyder2007} and the development of photonic spin switches~\cite{Lagoudakis2002,Amo2010a}. While the advantages of spintronics for information processing remain promising, the emerging field of valleytronics proposes to encode information in the valley degree of freedom of multivalley semiconductors~\cite{Behnia2012,Nebel2013}, including transition metal dichalcogenides~\cite{Xu2014,Jones2014,He2014,Chernikov2014}. This raises the question of whether valleytronics is itself a universal concept that can also appear in suitably engineered photonic systems.

A few recent works have taken an approach to hybridize light confined in planar microcavities with transition metal dichalcogenides~\cite{Vasilevskiy2015,Lundt2016} resulting in exciton-polaritons (EPs) with large binding energy. Indeed, this system is highly promising as a nonlinear photonic system operating at room temperature, however, the valleytronic features of multivalley semiconductors that occur at wave vectors given by the inverse crystal lattice constant are uncoupled to optical modes that are restricted to lower in-plane wave vectors inside the light cone. The engineering of multiple valleys suitable for nonlinear optical valleytronics thus requires a different approach.

Nevertheless, EPs remain a good candidate as their relatively large micron-scale de Broglie wavelength does present the advantage that EPs can be strongly manipulated by micron scale potentials. Such potentials may be achieved either through spatial modulation of the photon energy~\cite{IdrissiKaitouni2006,Lai2007} or the exciton energy~\cite{Balili2007,Amo2010,Assmann2012,Cristofolini2013,Askitopoulos2013}. Periodic potential arrays have been introduced~\cite{Lai2007,CerdaMendez2010}, with different lattice geometries~\cite{Kim2013,CerdaMendez2013,Winkler2016}, leading to gap solitons~\cite{Ostrovskaya2013,Tanese2013}, flatbands~\cite{Jacqmin2014}, and Bloch oscillations \cite{FlayacBO1,FlayacBO2}. They also lead to new devices \cite{Router,RouterExp} and (theoretically) non-trivial topological properties~\cite{Karzig2015,Nalitov2015,Bardyn2015,Bardyn2016}.

In this Report we consider the behavior of EPs in a microcavity where both the optical and excitonic components are separately manipulated by a periodic potential. While not considered before, this could be achieved by ``proton implantation'' \cite{Schneider2015} in which the properties of quantum wells and semiconductor microcavities can be spatially patterned after growth. A different localization of photons and excitons by their respective potentials theoretically allows for a peculiar overlap of their wave functions that depends on the in-plane momentum. Remarkably, we yield the momentum-dependent coupling between excitons and photons, which gives rise to the formation of unusual dispersions with degenerate ground states at non-zero momenta, at the bottom of different valleys in reciprocal space.

We further show that different valleys have different polarizations, in analogy to the spin-valley coupling that forms the basis of valleytronics in two-dimensional (2D) semiconductor systems. For additional effects that arise from the unusual EP dispersion in our system, we consider the behaviour under non-resonant and resonant excitation conditions. In the former case, it is known that EPs may undergo a Bose-Einstein condensation~\cite{Kasprzak2006}, characterized by the breaking of U(1) phase symmetry and the appearance of a macroscopic coherent low-energy state. Other symmetries may also be broken during Bose-Einstein condensation, both in EP systems and other systems, including spin symmetry breaking~\cite{Sadler2006,Ohadi2012} translational symmetry breaking~\cite{Kanamoto2003}, and angular momentum symmetry breaking~\cite{Butts1999,Liu2015}. In our system, we find that there is also a spontaneous breaking of linear momentum symmetry, unprecedented in other systems, where the condensate may spontaneously choose between different valleys in the dispersion. Finally, we show that under resonant excitation, the presence of a two-mode squeezing due to polariton-polariton interactions leads to the onset of non-classical quantum correlations.\\

\section*{Dispersion of exciton-polaritons in a lattice}
\begin{figure}[!tb]
\centering
	\includegraphics[width=0.99\linewidth]{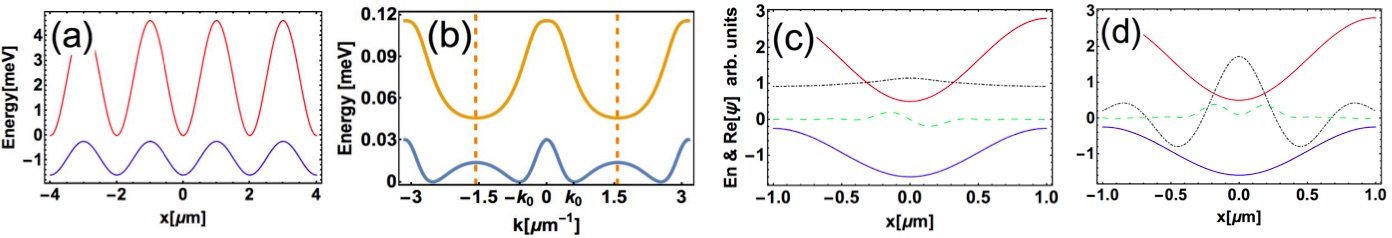}
	\caption{ System schematic. (a) Potential profiles for QW excitons (red) and cavity photons (blue).  (b) Numerically calculated dispersion of the lowest-energy EPs (blue). Orange {dashed} vertical lines label the edges of the first Brillouin zone. Yellow curve depicts the imaginary part of the polariton energy as a function of $k$. (c)\&(d) Real part of the wave function of cavity photons (black dot-dashed) and excitons (green dashed) in a single lattice period with $k=0$ (c) and $k$ at the edge of the Brillouin zone (d).}
	\label{Fig1}
\end{figure}

We  begin by considering a one-dimensional (1D) system of cavity photons and quantum-well (QW) excitons~\cite{Tanese2013}, which experience potentials with the same periodicity but different alignment in energy, as shown in Fig.~\ref{Fig1}a. Applying the Bloch theory and the model of coupled harmonic oscillators, we can use the central equation and solve the eigenvalue problem of the system, which in a brief form reads:
%
\begin{eqnarray}\label{SP1}
	\begin{pmatrix}
		\lambda_{C}-i\hbar/\tau_C-E & \Omega \\
		\Omega & \lambda_{X}-i\hbar/ \tau_X -E
	\end{pmatrix}
	C_k
	 + \sum_{G}
	\begin{pmatrix}
		\tilde{V}_{C}\left( G \right) & 0 \\
		0 & \tilde{V}_X\left( G \right)
	\end{pmatrix}
	C_{k-G}=0,
\end{eqnarray}
%
where $\lambda_{C}=\frac{\hbar^{2}k^{2}}{2m_{C}}$ and $\lambda_{X}=\frac{\hbar^{2}k^{2}}{2m_{X}}$ are the kinetic energy terms of the photonic and excitonic counterparts, correspondingly. Parameters $\tau_{C,X}$ are the lifetimes of the cavity photons and excitons, $\Omega$ is the exciton-photon coupling constant, $\tilde{V}_C\left( G \right)$ and $\tilde{V}_{X}\left( G \right)$ are the Fourier series coefficients of the potentials localizing the cavity photons and excitons in real space. The summation is over $G$, which is the reciprocal lattice vector of the periodic potential; $C_{k}$ are the (vector) amplitudes of the wave functions of polaritons in the photon-exciton basis with various $k$; $E$ are the eigenenergies of the EP modes.

After solution of the eigenvalue problem (see details in Supplementary), we find the dispersion of the system of EPs in $k$-space shown in Fig.~\ref{Fig1}b and the wave functions of photons and excitons (shown in Fig.~\ref{Fig1}c,d). It should be noted that the excitonic dispersion is nearly flat on the inverse $\mu m$ scale, such that excitons are localized inside the minima of their potential.

In the mean time, photons can also be localized depending on their momentum, which makes the overlap of photons and excitons momentum-dependent.
As it follows from Fig.~\ref{Fig1}b, the dispersion is characterized by two minima at non-zero wave vectors, $k=k_0$. This result is the milestone of this manuscript and in the following we show that this peculiarity leads to non-trivial effects such as spontaneous momentum symmetry breaking upon EP condensation and quantum entanglement.\\


\section*{Polariton condensation in the limit of thermal equilibrium}

Before considering the structure of the dispersion in 2D lattices, it is instructive to study the potential consequences of the dispersion shown in Fig.~\ref{Fig1}b for the 1D case. Here we begin by considering the behavior of the system under non-resonant excitation under which polariton condensation can be expected in the lattice~\cite{Lai2007}. Due to their finite lifetime, exciton-polaritons are non-equilibrium systems and so would not necessarily form in the ground state~\cite{Maragkou2010}, however, at high densities energy relaxation is typically enhanced to the ground state~\cite{Winkler2016}. In this section, we consider qualitative arguments in the limit of thermal equilibrium~\cite{Sun2017}. This is only intended to be used as a qualitative insight for more accurate non-equilibrium modeling that will be presented later in section~\ref{sec:Nonequilibrium}.

Given the dispersion, $E_k$, obtained in the linear regime (shown in Fig.~\ref{Fig1}b), the Hamiltonian of the system can be written as
\begin{equation}
	\label{H-2}
	\hat{\mathcal{H}}=\sum_k  E_k \hat{a}_k^\dag \hat{a}_k +\alpha \hat{a}_k^\dag \hat{a}_k^\dag \hat{a}_k \hat{a}_k + 2\alpha \sum_{k' \neq k} \hat{a}_k^\dag \hat{a}_{k'}^\dag \hat{a}_k \hat{a}_{k'},
\end{equation}
where we introduce polariton-polariton interaction with the strength $\alpha$. The factor $2$ in Eq.\eqref{H-2}  is characteristic of the momentum space scattering processes~\cite{Whittaker2005} and can be related to  inequivalent permutations of $\hat{a}_k^\dag \hat{a}_{k'}^\dag \hat{a}_k \hat{a}_{k'}$.
\begin{figure}[tb!]
	\centering
	\includegraphics[width=0.5\linewidth]{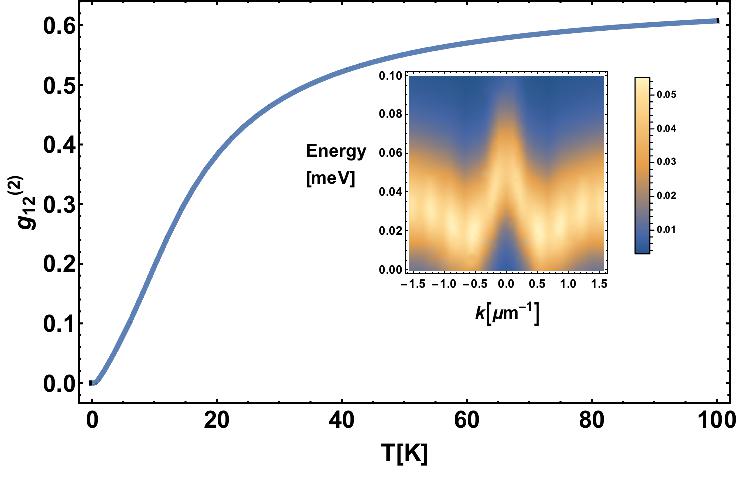}
	\caption{Second-order correlation function as a function of temperature. The total number of EPs is fixed, $n=100$. Inset picture demonstrates the low-density result with the total number of polaritons restricted to $n=3$, compare with the dispersion in Fig.~\ref{Fig1}b.}
	\label{TMFig}
\end{figure}

At zero temperature, one can expect that only the two lowest energy momentum states at $k_1= -k_0$ and $k_2=k_0$ are populated, where $k_0$ is the momentum of the right minimum of the blue curve in Fig.~\ref{Fig1}b. Then the energy of the system can be written as
\begin{align}
	E\left( n_1 , n_2 \right) &= nE_{k_1} + \alpha \left( n_1^2 +n_2^2 - n + 4n_1 n_2 \right),
	\label{E-2}
\end{align}
where for simplicity we denoted $n_{k_1} = n_1 $, $n_{k_2} = n_2$ and the total population $n =n_1 +n_2$.
The lowest energy state thus appears when $\rho=\left( n_1-n_2 \right)/n$ achieves its extreme value of $\pm 1$. In other words, at zero temperature we expect the system to spontaneously choose either the state with all the EPs at $k_1$ or $k_2$. 
This can be further confirmed by calculating the second order correlation function and spectrum corresponding to Hamiltonian \ref{H-2}, as shown in Fig.~\ref{TMFig} (details of the calculation are included in Supplementary).\\


\section*{Nonequilibrium model of polariton condensation \label{sec:Nonequilibrium}}

EPs have finite lifetime and consequently form non-equilibrium condensates. In samples with weak energy relaxation, they do not necessarily reach the actual ground state of the system~\cite{Krizhanovskii2009, Maragkou2010}.
Let us further investigate the behavior of the system using a stochastic quantum treatment and accounting for various scattering processes (see Supplementary).
We considered an InGaAlAs alloy-based microcavity and in computations used the following parameters: speed of sound $c_s=5370$ $m/s$~\cite{Hartwell2010}, $\gamma=i\hbar/\tau=\hbar/18$ $ps^{-1}$.~\cite{Gao2012}

\begin{figure}[!t]
\centering
	\includegraphics[width=0.99\linewidth]{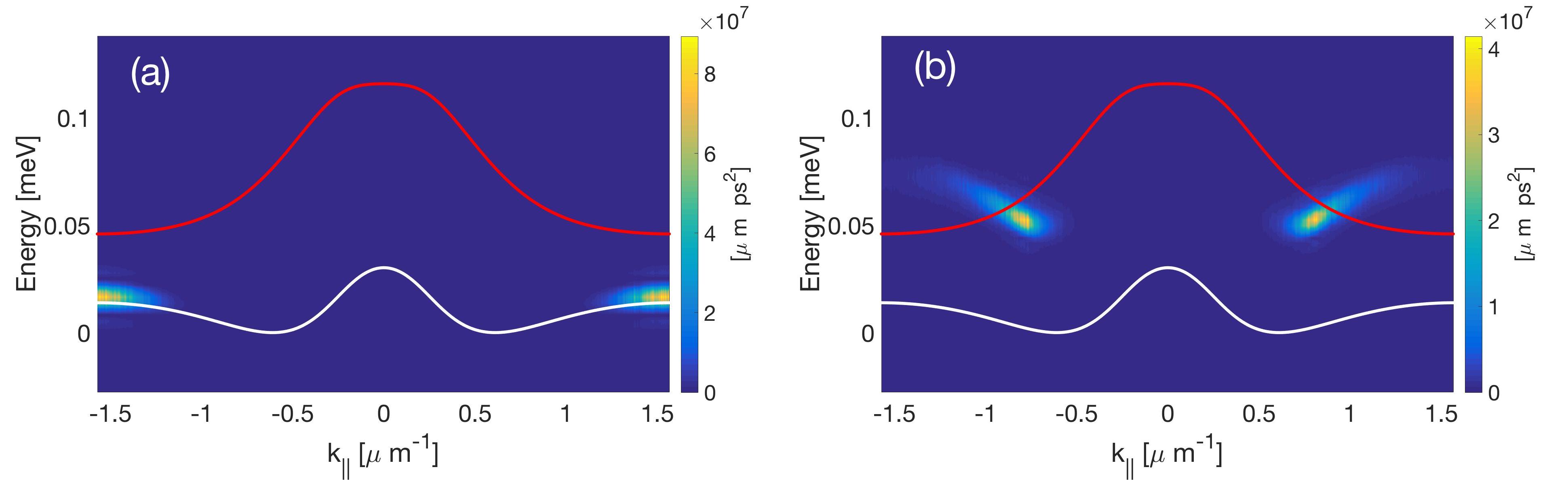}
	\caption{Distribution of EPs localized in the potential profile shown in Fig.~\ref{Fig1}b in the regime of homogeneous incoherent excitation of the system in the steady state (at 1500 ps) with account of the acoustic phonon-assisted scattering.
	Red curves show the $k$-dependence of the decay rates. White curves show the EP dispersion in the linear regime.
	EP condensates form in the reciprocal space at $k_\parallel\approx \pm 0.74$ $\mu m^{-1}$.
The polariton-polariton interaction is switched off (a) and on (b) by putting $\alpha$ zero and non-zero, respectively.}
\label{Fig4}
\end{figure}

In Fig. \ref{Fig4}a, we switch off the polariton-polariton interaction and see that in this case there is no blueshift and the particles occupy mostly the edge of the Brillouin zone. It happens due to the fact that the lifetime of particles increases with the increase of $|k_{\parallel}|$ and thus the decay rate decreases with $|k_{\parallel}|$, see red curve in Fig.~\ref{Fig4}.
However, with account of the interaction, we achieve the degenerate condensation at points $k=\pm k_0$ due to the interplay of particle lifetime and interactions, see Fig.~\ref{Fig4}b. EPs are blueshifted in energy (compare with Fig.~\ref{Fig4}a).

It is important to note, that if we change the potential profiles for the excitons and photons (change the shapes of the curves in Fig.~\ref{Fig1}a), we can achieve different points of condensation, in particular we can make particles condense at $k=0$ and $k=k_{BZ}$, see Supplementary.

\begin{figure}[b!]
\centering
	\includegraphics[width=0.8\linewidth]{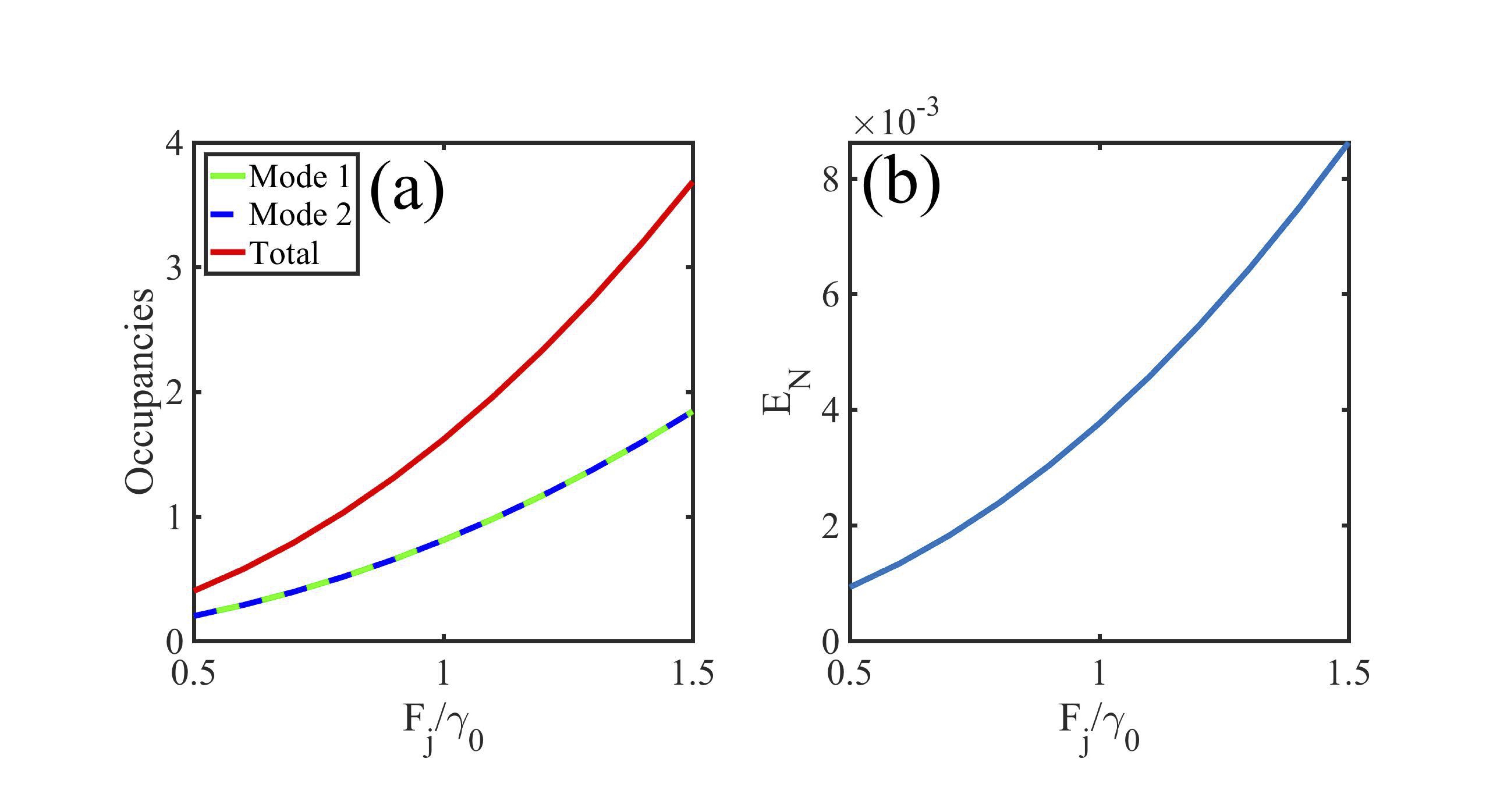}
	\caption{Entanglement generation between the particles localized in the dispersion minima under resonant excitation (see Fig.~1a,c and d). (a) Average mode occupancies found from the numerical solution of Eq.~\eqref{rhot} in the steady state and (b) logarithmic negativity versus increasing driving strength. The parameters are $\gamma_0=8.6$ ps$^{-1}$, $\Delta_1=\Delta_2=\hbar\gamma_0$, $\alpha=4.5\times10^{-4} {\rm{meV}} $.  }
	\label{Entanglement}
\end{figure}

While the condensation of EPs to non-zero momentum states has been observed previously~\cite{Liu2015,Maragkou2010,Kusudo2013} in those observations it was a purely non-equilibrium effect. In our work, the condensation to non-zero momentum takes place even in  the limit of equilibrium, that is, strong energy relaxation. Furthermore, since the non-zero momentum states represent the true ground-state of the system, they are likely to be highly stable after they have formed, particularly in polariton systems close to thermal equilibrium. This may include recently developed long-lifetime inorganic microcavities~\cite{Nelson2013} as well as organic systems~\cite{KenaCohen2010} with faster energy relaxation processes~\cite{Lanty2008}.\\


\section*{Entanglement generation: Resonant Excitation}

The two degenerate dispersion minima within the first Brillouin zone (shown in Fig.~1b and also in Fig.~2) offer a unique opportunity to study controlled entanglement in the system. Indeed, let us assume that each minimum is driven by a cw laser with large enough spatial extension thus we can consider only two quantum modes, described by the creation operators $\hat a_{1}^{\dag}$ and $\hat a_{2}^{\dag}$, respectively.

\begin{figure}[!tb]
	\includegraphics[width=0.99\linewidth]{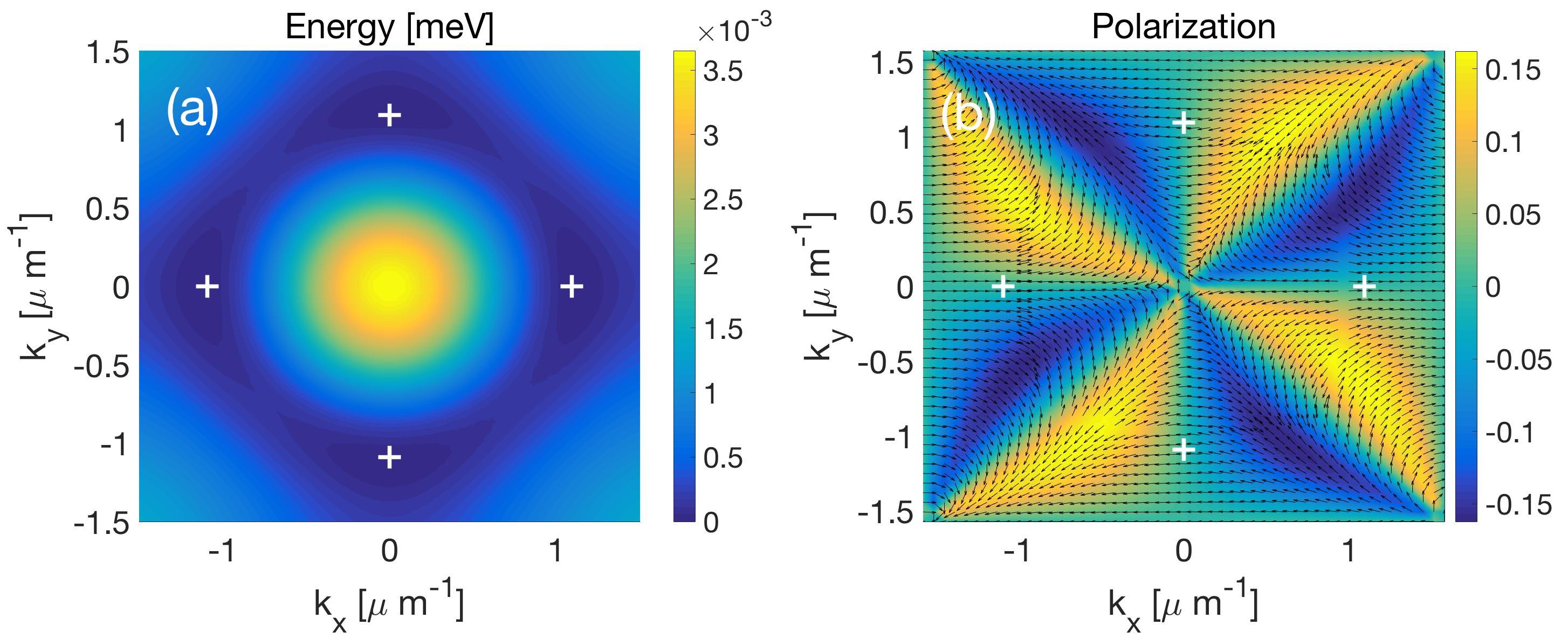}
\caption{Illustration of the multivalley coupling. (a) the energy dispersion in 2D, the picture shows the result for the First Brillouin zone. (b) polarization chart. The arrows show the polarization in $x$ and $y$ directions, the color represent the polarization in $z$ direction. White crosses correspond to the minima of the energy dispersion.}
\label{Fig2D}
\end{figure}

In order to describe the dynamics of such system, we use a dissipative master equation in the form:
\begin{equation}
	\label{rhot}
	i\hbar\frac{{\partial \hat \rho }}{{\partial t}} = [ {\hat {\cal H}',\hat \rho } ] - \frac{{i{\hbar\gamma _0}}}{2}\sum\limits_{j = 1,2} {\hat {\cal D}\left[ {{{\hat a}_j}} \right]\hat \rho },
\end{equation}
where the deterministic evolution (the first term on the rhs) is described by the Hamiltonian,
\begin{eqnarray}
	 {\cal \hat H}' &=& \sum\limits_{j = 1,2} {-{\Delta _j}\hat a_j^\dag {{\hat a}_j} + \alpha\hat a_j^\dag \hat a_j^\dag {{\hat a}_j}{{\hat a}_j}}  + {F_j}\left( {\hat a_j^\dag  + {{\hat a}_j}} \right) +4\alpha\hat a_1^\dag \hat a_2^\dag {{\hat a}_2}{{\hat a}_1},
		\label{Hres}
\end{eqnarray}
which is similar to \eqref{H-2} after adding the pumping terms and in the frame rotating with the laser's frequency, assuming $F_j\in\mathbb{R}$ and with $\Delta_j$ defined as the laser/modes detuning; let us also assume $D_1=D_2=0$ and $F_1=F_2$ for simplicity.
It is crucial that the Hamiltonian~\eqref{Hres} includes two kinds of interaction terms. The first one ($\alpha$) stands for the interaction of EPs located within the same minimum, while the second term ($4\alpha$) accounts for the cross-interactions between the minima which fulfills the energy momentum conservation by exchange of the wave vector $+k\rightarrow-k$ and no energy change (see Supplementary for details). Physically, the latter term corresponds to the cross-Kerr interaction which is known to induce two-mode squeezing~\cite{CrossKerr}, and therefore continuous variable entanglement is expected to occur in reciprocal space. %

In the indeterministic part of the evolution described by the last term in~\eqref{rhot}, $\hat{\cal{D}}\left[ {{{\hat a_j}}} \right]\hat \rho = \{\hat a_j^\dag {{\hat a_j}},\hat \rho\} - 2{{\hat a_j}}\hat \rho \hat a_j^\dag$ correspond to the Lindblad superoperators which account for polariton losses due to the interaction with their environment. The continuous variable entanglement is quantified via the \emph{logarithmic negativity} \cite{Vidal2002}, $E_{\cal N}=\log_2{|| {\hat \rho^{{\Gamma _{2}}}} ||_1}$, where ${\hat \rho^{{\Gamma _{2}}}}$ is the partial transpose of $\hat \rho$ with respect to the second mode and ${|| {\hat \rho^{{\Gamma _{2}}}} ||_1}$ is its trace norm. In Fig.~\ref{Entanglement} we present steady state numerical solutions to Eq.~\eqref{rhot} showing the average mode occupancies in panel (a) and the corresponding values of $E_{\cal N}$ in panel (b) for increasing amplitude of the pumps.
We observe a monotonous increase of both the quantities, as a direct proof of entanglement. Importantly, no state other than the one targeted by the laser can be reached by potential parametric processes which would violate the energy conservation.\\


\section*{Polarization-Valley coupling in 2D}

Finally, we present the calculation of the dispersion in a 2D system obtained from generalization of the result in 1D to a square lattice. Fig.~\ref{Fig2D}a shows the energy of the system ground state in the first Brillouin zone (see also the 3D plot in Supplementary). Here we can identify four energy minima, at the bottom of different valleys in reciprocal space.

A fundamental feature of 2D semiconductors for valleytronics is the spin-valley coupling that allows different valleys to be excited with light of different polarization~\cite{Xu2014,Jones2014,He2014,Chernikov2014}. In the system of exciton-polaritons, it is well-known that transverse-electric and transverse-magnetic polarized modes are split in energy. This splitting can be modeled by introducing a spin-orbit coupling Hamiltonian acting on the photon spin degree of freedom~\cite{Ohadi2012}:
\begin{equation}
\mathcal{H}_{TE-TM}=\left(\begin{array}{cc}0&\Delta\left(i\frac{\partial}{\partial x}+\frac{\partial}{\partial y}\right)^2\\\Delta\left(i\frac{\partial}{\partial x}-\frac{\partial}{\partial y}\right)^2&0\end{array}\right).
\end{equation}

Accounting for this splitting, we obtain the polarization structure of the lowest energy band shown in Fig.~\ref{Fig2D}b. Here we note that different valleys have different polarizations, which implies that they can be selectively excited by a resonant excitation of specific polarization.
The geometry of the lowest energy pseudospin should allow to form such patterns as skyrmions~\cite{Hugo2013,Cilibrizzi2016} or spin whirls~\cite{Cilibrizzi2015} under pulsed-resonant excitation.\\

\section*{Conclusion}

We have considered the formation of exciton-polaritons in a semiconductor microcavity with separate spatially patterned potentials for cavity photons and excitons. The different confinement of photons and excitons allows for a momentum dependent coupling which gives rise to a unique form of the  dispersion in which degenerate ground states appear at non-zero momenta. We considered two different limits corresponding to strong and weak energy relaxation. In the limit of strong energy relaxation, a simple equilibrium theoretical model predicts spontaneous symmetry breaking in momentum space. In the limit of weak energy relaxation, a non-equilibrium model accounting for phonon scattering processes shows non-equilibrium condensation at non-zero wave vector. Treating exciton polaritons as an open quantum system, we have also shown that correlations between the modes in reciprocal space can occur. Finally, considering exciton-polaritons in a 2D square lattice, we predict the formation of a multivalley-dispersion. Here different valleys exhibit different polarizations, which, in principal, allows for their selective excitation by a polarized laser and forms a foundation for exciton-polariton valleytronics.



\section*{Acknowledgements}

We thank Prof. Sergej Flach for fruitful discussions. M. Sun thanks Dr. Pinquan Qin for the discussions personally.
We acknowledge support of Project Code (IBS-R024-D1); the Australian Research Council's Discovery Projects funding scheme (project DE160100167), the government of Russian Federation (project MK-5903.2016.2) and the Dynasty Foundation. TCHL was supported by the MOE AcRF Tier 1 grant 2016-T1-1-084 and MOE
AcRF Tier 2 grant 2015-T2-1-055.\\

\section*{Author contributions statement}
MS performed the calculation of the dispersion, the distributions of particles under equilibrium and nonequilibrium conditions, and the spin-valley coupling under the supervision of IGS. TCHL conceived the project where HF conceived and calculated entanglement using a quantum optical model. All authors contributed to the analysis of results and writing of the manuscript.

\section*{Competing interests}
The authors declare no competing financial interests.



\begin{thebibliography}{1}	

    \bibitem{Lu2014}
    Lu, L., Joannopoulos, J. D. \& Solja\u{c}i\'{c}, M. Topological photonics. \textit{Nat. Photon.} {\bf 8}, 821 (2014).

    \bibitem{Leyder2007}
    Leyder, C., et al. Observation of the optical spin Hall effect. \textit{Nature Phys.} {\bf 3}, 628 (2007).

    \bibitem{Lagoudakis2002}
    Lagoudakis. P. G., et al. Stimulated spin dynamics of polaritons in semiconductor microcavities. \textit{Phys. Rev. B} {\bf 65} 161310(R)
    
    \bibitem{Amo2010a}
    Amo, A., et al. Exciton-polariton spin switches. \textit{Nat. Photon.} {\bf 4}, 361 (2010).
    
        \bibitem{Behnia2012}
    Behnia, K. Condensed-matter physics: Polarized light boosts valleytronics. \textit{Nat. Nanotechnol.} {\bf 7}, 488 (2012).

    \bibitem{Nebel2013}
    Nebel, C. E. Valleytronics: Electrons dance in diamond. \textit{Nat. Mater.} {\bf 12}, 690 (2013).

    \bibitem{Xu2014}
    Xu, X., Yao W., Xiao, D. \& Heinz, T. F. Spin and pseudospins in layered transition metal dichalcogenides. \textit{Nature Phys.} {\bf 10}, 343 (2014).

    \bibitem{Jones2014}
    Jones, A. M., et al. Spin-layer locking effects in optical orientation of exciton spin in bilayer $WSe_2$. \textit{Nature Phys.} {\bf 10}, 130 (2014).

    \bibitem{He2014}
    He, K., et al. Tightly bound excitons in monolayer $WSe_2$. \textit{Phys. Rev. Lett.} {\bf 113}, 026803 (2014).

    \bibitem{Chernikov2014}
    Chernikov, A., et al. Exciton binding energy and nonhydrogenic Rydberg series in monolayer $WS_2$. \textit{Phys. Rev. Lett.} {\bf 113}, 076802 (2014).

    \bibitem{Vasilevskiy2015}
   Vasilevskiy, M. I., et al. Exciton polaritons in two-dimensional dichalcogenide layers placed in a planar microcavity: Tunable interaction between two Bose-Einstein condensates. \textit{Phys. Rev. B} {\bf 92}, 245435 (2015).

    \bibitem{Lundt2016}
    Lundt, N., et al. Monolayered $MoSe_2$: a candidate for room temperature polaritonics. \textit{2D Mater.} {\bf 4}, 015006 (2016).

	\bibitem{IdrissiKaitouni2006}
		Kaitouni, R. I., et al. Engineering the spatial confinement of exciton polaritons in semiconductors. \textit{Phys. Rev. B} {\bf 74}, 155311 (2006).

	\bibitem{Lai2007}
		 Lai, C. W., et al. Coherent zero-state and $\pi$-state in an exciton-polariton condensate array. \textit{Nature} {\bf 450}, 529 (2007).
		 
\bibitem{Balili2007}
		Balili, R., Hartwell, V., Snoke, D. Pfeiffer, L. \& West, K. Bose-Einstein condensation of microcavity polaritons in a trap. \textit{Science} {\bf 316}, 1007 (2007).

	\bibitem{Amo2010}
		Amo, A., et al. Light engineering of the polariton landscape in semiconductor microcavities. \textit{Phys. Rev. B} {\bf 82}, 081301R (2010).

	\bibitem{Assmann2012}
		A\ss mann, M., et al. All-optical control of quantized momenta on a polariton. \textit{Phys. Rev. B} {\bf 85}, 155320 (2012).

	\bibitem{Cristofolini2013}
		Cristofolini, P., et al. Optical superfluid phase transitions and trapping of polariton condensates. \textit{Phys. Rev. Lett.} {\bf 110}, 186403 (2013).

	\bibitem{Askitopoulos2013}
		Askitopoulos, A., Ohadi, H., Kavokin, A. V., Hatzopoulos, Z., Savvidis, P. G. \& Lagoudakis, P. G. Polariton condensation in an optically induced two-dimensional potential. \textit{Phys. Rev. B} {\bf 88}, 041308(R) (2012).

\bibitem{CerdaMendez2010}
		Cerda-M\'endez, E. A., et al. Polariton condensation in dynamic acoustic lattices. \textit{Phys. Rev. Lett.} {\bf 105}, 116402 (2010).

\bibitem{Kim2013}
		Kim, N. Y., Kusudo, K., L\"offler, A., H\"ofling,  S., Forchel, A. \& Yamamoto, Y. Exciton-polariton condensates near the Dirac point in a triangular lattice. \textit{New J. Phys.} {\bf 15}, 035032 (2013).

	\bibitem{CerdaMendez2013}
		 Cerda-M\'endez, E. A., et al. Exciton-polariton gap solitons in two-dimensional lattices. \textit{Phys. Rev. Lett.} {\bf 111}, 146401 (2013).

	\bibitem{Winkler2016}
		 Winkler, K., et al. Collective state transitions of exciton-polaritons loaded into a periodic potential. \textit{Phys. Rev. B} {\bf 93}, 121303(R) (2016).

	\bibitem{Ostrovskaya2013}
		Ostrovskaya, E. A., Abdullaev, J., Fraser, M. D., Desyatnikov, A. S. \&  Kivshar, Yu. S. Self-localization of polariton condensates in periodic potentials. \textit{Phys. Rev. Lett.} {\bf 110}, 170407 (2013).

	\bibitem{Tanese2013}
		Tanese, D., et al. Polariton condensation in solitonic gap states in a one-dimensional periodic potential. \textit{Nature Comm.} {\bf 4}, 1749 (2013).

	\bibitem{Jacqmin2014}
		Jacqmin, T., et al. Direct observation of Dirac cones and a flatband in a honeycomb lattce for polaritons. \textit{Phys. Rev. Lett.} {\bf 112}, 116402 (2014).

	\bibitem{FlayacBO1} 
	Flayac, H., Solnyshkov, D. D. \& Malpuech, G. Bloch oscillations of an exciton-polariton Bose-Einstein condensate. \textit{Phys. Rev. B} {\bf 83}, 045412 (2011).

	\bibitem{FlayacBO2} 
	Flayac, H., Solnyshkov, D. D. \& Malpuech, G. Bloch oscillations of exciton-polaritons and photons for the generation of an alternating terahertz spin signal. \textit{Phys. Rev. B} {\bf 84}, 125314 (2011).

	\bibitem{Router} 
	Flayac, H. \& Savenko, I. G. An exciton-polariton mediated all-optical router.  \textit{Appl. Phys. Lett.} {\bf 103}, 201105 (2013).

	\bibitem{RouterExp} 
	Marsault, F., et al. Realization of an all optical exciton-polariton router. \textit{Appl. Phys. Lett.} {\bf 107}, 201115 (2015).

	\bibitem{Karzig2015}
		Karzig, T., Bardyn, C-E., Lindner, N. H. \& Refael, G. Topological polaritons. \textit{Phys. Rev. X} {\bf 5}, 031001 (2015).

	\bibitem{Nalitov2015}
		Nalitov, A. V., Solnyshkov, D. D. \& Malpuech, G. Polariton Z topological insulator. \textit{Phys. Rev. Lett.} {\bf 114}, 116401 (2015).

	\bibitem{Bardyn2015}
		Bardyn, C-E., Karzig, T., Refael, G. \& Liew, T. C. H. Topological polaritons and excitons in garden-variety systems. \textit{Phys. Rev. B} {\bf 91}, 161413(R) (2015).

	\bibitem{Bardyn2016}
		Bardyn, C-E., Karzig, T., Refael, G. \& Liew, T. C. H. Chiral Bogoliubov excitations in nonlinear bosonic systems. \textit{Phys. Rev. B} {\bf 93}, 020502(R) (2016).

\bibitem{Schneider2015}
		Schneider, C., et al. Exciton-polariton trapping and potential landscape engineering. arXiV: 1510.07540 (2015).

\bibitem{Kasprzak2006} 
	Kasprzak, J., et al. Bose-Einstein condensation of exciton polaritons. \textit{Nature} \textbf{443}, 409 (2006).

\bibitem{Sadler2006}
		Sadler, L. E., Higbie, J. M., Leslie, S. R., Vengalattore, M. \& Stamper-Kurn, D. M. Spontaneous symmetry breaking in a quenched ferromagnetic spinor Bose-Einstein condensate. \textit{Nature}, {\bf 443}, 312 (2006).

\bibitem{Ohadi2012}
		Ohadi, H., Kammann, E., Liew, T. C. H., Lagoudakis, K. G., Kavokin, A. V. \& Lagoudakis, P. G. Spontaneous symmetry breaking in a polariton and photon laser. \textit{Phys. Rev. Lett.} {\bf 109}, 016404 (2012).

\bibitem{Kanamoto2003}
		Kanamoto, R., Saito, H. \& Ueda, M. Quantum phase transition in one-dimensional Bose-Einstein condensates with attractive interactions. \textit{Phys. Rev. A} {\bf 67}, 013608 (2003).

\bibitem{Butts1999}
		Butts, D. A. \& Rokhsar, D. S. Predicted signatures of rotating Bose-Einstein condensates. \textit{Nature}, {\bf 397}, 327 (1999).

\bibitem{Liu2015}
		Liu, G., Snoke, D. W., Daley, A. Pfeiffer, L. N. \& West, K. A new type of half-quantum circulation in a macroscopic polariton spinor ring condensate. \textit{Proc. Natl. Acad. Sci. U.S.A.} {\bf 112}, 2676 (2015).

\bibitem{Maragkou2010}
		Maragkou, M., et al. Spontaneous nonground state polariton condensation in pillar microcavities. \textit{Phys. Rev. B} \textbf{81}, 081307(R) (2010).

\bibitem{Kusudo2013}
		Kusudo, K., et al. Stochastic formation of polariton condensates in two degenerate orbital states. \textit{Phys. Rev. B.} {\bf87}, 214503 (2013).


\bibitem{Sun2017}
Sun Yongbao, et al. Bose-Einstein condensation of long-life polaritons in thermal equilibrium. \textit{Phys. Rev. Lett.} {\bf 118}, 016602 (2017).

\bibitem{Whittaker2005}
		Whittaker, D. M. Effects of polariton-energy renormalization in the microcavity optical parametric oscillator. \textit{Phys. Rev. B} {\bf 71}, 115301 (2005).

\bibitem{Krizhanovskii2009}
		Krizhanovskii, D. N., et al. Coexisting nonequilibrium condensates with long-range spatial coherence in semiconductor microcavities.	 \textit{Phys. Rev. B} \textbf{80}, 045317 (2009).

\bibitem{Hartwell2010}
		Hartwell, V. E.  \& Snoke, D. W. Numerical simulations of the polariton kinetic energy distribution in GaAs quantum-well microcavity structures. \textit{Phys. Rev. B} \textbf{82}, 075307 (2010).

	\bibitem{Gao2012} 
	Gao, T., et al. Polariton condensate transistor switch. \textit{Phys. Rev. B} \textbf{85} 235102 (2012).

\bibitem{Nelson2013}
		Nelsen, B., et al. Dissipationless flow and sharp threshold of a polariton condensate with long lifetime. \textit{Phys. Rev. X} {\bf 3}, 041015 (2013).

	\bibitem{KenaCohen2010}
		K\'ena-Cohen, S. \& Forrest, S. R. Room-temperature polariton lasing in an organic single-crystal microcavity. \textit{Nat. Photon.} {\bf 4}, 371-375 (2010).

	\bibitem{Lanty2008}
		Lanty, G., Lauret, J. S., Deleporte, E., Bouchoule, S. \& Lafosse, X. UV polaritonic emission from a perovskite-based microcavity. \textit{Appl. Phys. Lett.} {\bf 93} 081101 (2008).

\bibitem{CrossKerr}
	Wang, T., Lau, H. W., Kaviani, H., Ghobadi, R. \& Simon, Christoph. Strong micro-macro entanglement from a wark cross-Kerr nonlinearity. \textit{Phys. Rev. A } {\bf 92}, 012316 (2015).

	 \bibitem{Vidal2002}
		 Vidal, G. \& Werner, R. F. Computable measure of entanglement. \textit{Phys. Rev. A}  {\bf 65}, 032314 (2002).
		 \bibitem{Hugo2013}
		 Flayac, H. et al. Transmutation of skyrmions to half-solitons driven by the nonlinear optical spin Hall effect. \textit{Phys. Rev. Lett.} {\bf 110}, 016404 (2013)
		 \bibitem{Cilibrizzi2016}
		 Cilibrizzi, P., et al. Half-skyrmion spin texture in polariton microcavities. \textit{Phys. Rev. B}  {\bf 94}, 045315 (2016)
		 
		 \bibitem{Cilibrizzi2015}
		 Cilibrizzi, P., et al. Polariton spin whirls. \textit{Phys. Rev. B} {\bf 92} 155308 (2015)


%
%
%
%
%
%
%
%
%
%
%
%
%
%
%
%
%
%
%
%
%
%
%
%
%
%
%
%
%
%
%
%
%
%
%
%
%
%
%
%
%
%
%
%
%
%


\end{thebibliography}
\end{document}